\documentclass[preprint*]{JHEP3} 

\usepackage{epsfig,multicol}

\newcommand\fverb{\setbox\pippobox=\hbox\bgroup\verb}
\newcommand\fverbdo{\egroup\medskip\noindent%
                        \fbox{\unhbox\pippobox}\ }
\newcommand\fverbit{\egroup\item[\fbox{\unhbox\pippobox}]}
\newbox\pippobox

\newcommand{\beq}{\begin{equation}}
\newcommand{\eeq}{\end{equation}}
\newcommand{\nn}{\nonumber}
\newcommand{\bea}{\begin{eqnarray}}
\newcommand{\eea}{\end{eqnarray}}

\newcommand{\Eq}[1]{Eq.~(\ref{#1})}

\title{Finite $SU(N)^k$ Unification}

\author{Ernest Ma,$^1$\thanks{Supported in part by the
    U.~S.~Department of Energy under Grant No.~DE-FG03-94ER40837, } 
Myriam Mondrag{\'o}n$^2$\thanks{Supported in part by the mexican grants
  PAPIIT-UNAM 
IN116202, and Conacyt grant 42026-F}~
        and George Zoupanos$^3$\thanks{Partially
supported by the programmes of Ministry of Education
``$\Pi$Y$\Theta$A$\Gamma$OPA$\Sigma$ " and
``HPAK$\Lambda$EITO$\Sigma$ " and by the NTUA programme
``$\Theta$A$\Lambda$H$\Sigma$".}\\
        $^1$ Physics Department, University of California, Riverside,
California 92521, USA\\  $^2$ Inst.~de Fisica, Univ.~Nacional Autonoma
de Mexico, Mexico 01000, D.F., Mexico\\ 
$^3$ Physics Department, National Technical University of Athens, 
Athens, Greece \\ 
        E-mail: \email{},
        \email{}}  

\preprint{UCRHEP-T376\\ July 2004}      
      
\abstract{We consider $N=1$ supersymmetric gauge theories based on the
  group $SU(N)_1 \times SU(N)_2 \times ... \times SU(N)_k$ with matter content
  $(N,N^*,1,...,1) + (1,N,N^*,...,1) + ... + (N^*,1,1,...,N)$ as
  candidates for the unification symmetry of all particles.  In
  particular we examine to which extent such theories can become
  finite and we find that a necessary condition is that there should
  be exactly three families.  We discuss further some phenomenological
  issues related to the cases $(N,k) = (3,3)$, (3,4), and (4,3), in an
  attempt to choose those theories that can become also realistic.
  Thus we are naturally led to consider the $SU(3)^3$ model which we
  first promote to an all-loop finite theory and then we study its
  additional predictions concerning the top quark mass, Higgs mass and
  supersymmetric spectrum.}


\begin{document}

\section{Introduction}

Finite field theories \cite{lucchesi1,ermushev1,finite1,kkmz1,acta}
are very attractive since they are  free of all ultraviolet divergencies,
but require a large degree of symmetry, which obviously is not
observed at low energies.  However, the intriguing possibility exists
that the standard model (SM) we observe is a remnant of a finite Grand
Unified Theory (GUT) at the unification scale and above.  This may
provide the missing deep connection of current phenomenology with
string theory and may point to a unique candidate for the description
of all fundamental interactions.

Finite Unified Theories (FUTs) are $N=1$ supersymmetric GUTs, which
can be made finite even to all-loop orders, including the soft
supersymmetry breaking sector.  The method to construct GUTs with
reduced independent parameters \cite{zim1,kmz1} consists of searching
for renormalization group invariant (RGI) relations holding below the
Planck scale, which in turn are preserved down to the GUT scale. Of
particular interest is the possibility to find RGI relations among
couplings that guarantee finiteness to all-orders in perturbation
theory \cite{lucchesi1,ermushev1}. In order to achieve the latter it
is enough to study the uniqueness of the solutions to the one-loop
finiteness conditions \cite{lucchesi1,ermushev1,acta}. Using the above
tools elegant $N=1$ supersymmetric $SU(5)$ examples already exist, and
have predicted correctly from the dimensionless sector, among others,
the top quark mass \cite{finite1,kkmz1}.  The search for RGI relations
has been extended to the soft supersymmetry breaking sector (SSB) of
these theories \cite{kmz2,jack2}, which involves parameters of
dimension one and two.

Here we examine the construction of realistic FUTs based on product
gauge groups. In particular we point out that finiteness actually
determines the number of families $n_f$ in a class of supersymmetric
$SU(N)^k$ gauge theories, namely $n_f = 3$ regardless of $N$ and $k$.
The case $N=4$ and $k=3$ was first pointed in ref.~\cite{su4^3}, and
that of arbitrary $N$ and $k=3$ was discussed in ref.~\cite{sun^3},
both from the string point of view.  Concerning the soft supersymmetry
breaking sector of these latter models, although in principle it could
be understood too in the same framework under certain assumptions
\cite{su4^3,Brignole:1995fb,Camara:2003ku}, the explicit construction is
still missing.

Our search for realistic FUTs based on product groups leads us to
choose a supersymmetric $SU(3)^3$ model, which we subsequently promote
to an all-loop finite theory,  whose predictions we examine further.

The rest of the paper is organised as follows.  In section 2 we review
the method of reduction of couplings and recall how it is applied in
$N=1$ supersymmetric gauge theories in order to obtain all-loop finite
gauge theories.  In section 3 we describe the extension of finiteness
in the case of soft supersymmetry breaking terms.  Section 4 is
devoted to a search for realistic FUTs based on product groups, out of
which an $SU(3)^3$ supersymmetric gauge theory with three families is
singled out.  This theory then is further discussed in detail in
section 5.  Section 6 contains the predictions of the $SU(3)^3$ FUT
concerning the top quark mass, the Higgs boson masses and the
supersymmetric spectrum.

\section{Reduction of Couplings and Finiteness in $N=1$ Supersymmetric
  Gauge Theories}

Let us first recall the basic issues concerning {\em reduction of
couplings}, in the case of dimensionless couplings and {\em
finiteness} of $N=1$ supersymmtric theories. 

A RGI relation among couplings $g_i$,  
\beq
 {\cal F} (g_1,\cdots,g_N) ~=~0, 
\eeq
has to satisfy the partial differential equation 
\beq
\mu~ d {\cal F} /d \mu
~=~ \sum_{i=1}^{N} \,\beta_{i}\,\partial {\cal F} /\partial g_{i}~=~0,
\eeq
where $\beta_i$ is the $\beta$-function of $g_i$.  There exist ($N-1$)
independent ${\cal F}$'s, and finding the complete set of these solutions
is equivalent to solve the so-called reduction equations (REs) \cite{zim1},
\beq
\beta_{g} \,(d g_{i}/d g) =\beta_{i}~,~i=1,\cdots,N,
\eeq
 where $g$ and
$\beta_{g}$ are the primary coupling and its $\beta$-function.  Using
all the $(N-1)\,{\cal F}$'s to impose RGI relations, one can in principle
express all the couplings in terms of a single coupling $g$.  The
complete reduction, which formally preserves perturbative
renormalizability, can be achieved by demanding a power series
solution, whose uniqueness can be investigated at the one-loop level.
 
In order to discuss finiteness, it seems unavoidable that we should
consider supersymmetric gauge theories. 
Let us then consider a chiral, anomaly free,
$N=1$ globally supersymmetric
gauge theory based on a group G with gauge coupling
constant $g$. The
superpotential of the theory is given by
\beq
W= \frac{1}{2}\,m^{ij} \,\Phi_{i}\,\Phi_{j}+
\frac{1}{6}\,C^{ijk} \,\Phi_{i}\,\Phi_{j}\,\Phi_{k}~,
\label{supot}
\eeq
where $m^{ij}$ and $C^{ijk}$ are gauge invariant tensors and
the matter field $\Phi_{i}$ transforms
according to the irreducible representation  $R_{i}$
of the gauge group $G$. 
All the one-loop $\beta$-functions of the theory
vanish if  $\beta_g^{(1)}$ and all the anomalous dimensions of the
superfields $\gamma_i^{j(1)}$ vanish, i.e.
\beq
\sum _i \ell (R_i) = 3 C_2(G) \,,~
\frac{1}{2}C_{ipq} C^{jpq} = 2\delta _i^j g^2  C_2(R_i)~,
\label{2nd}
\eeq where $l(R_i)$ is the Dynkin index of $R_i$, and $C_2(G)$,
$C_2(R_i)$ are respectively the quadratic Casimir invariant of the
adjoint representation of $G$, and of the $R_i$ representation. A
natural question to ask is what happens at higher loop orders.
A very interesting result is that the conditions (\ref{2nd}) are
necessary and sufficient for finiteness at
the two-loop level \cite{soft,PW}.

The one- and two-loop finiteness conditions (\ref{2nd}) restrict
considerably the possible choices of the irreps.~$R_i$ for a given
group $G$ as well as the Yukawa couplings in the superpotential
(\ref{supot}).  Note in particular that the finiteness conditions
cannot be applied to the supersymmetric standard model (SSM), since
the presence of a $U(1)$ gauge group is incompatible with the
first of the conditions (\ref{2nd}), due to $C_2[U(1)]=0$.  This leads to the
expectation that finiteness should be attained at the grand unified
level only, the SSM being just the corresponding, low-energy,
effective theory.

The finiteness conditions impose relations between gauge and Yukawa
couplings.  Therefore, we have to guarantee that such relations
leading to a reduction of the couplings hold at any renormalization
point.  The necessary, but also sufficient, condition for this to
happen is to require that such relations are solutions to the
reduction equations (REs) to all orders.  Specifically there exists a
very interesting theorem \cite{lucchesi1} which guarantees the
vanishing of the $\beta$-functions to all orders in perturbation theory,
if we demand reduction of couplings, and that all the one-loop
anomalous dimensions of the matter field in the completely and
uniquely reduced theory vanish identically.

\section{Soft Supersymmetry Breaking in $N=1$ FUTS}

The above described method of reducing the dimensionless couplings has
been extended \cite{kmz2} to the soft supersymmetry breaking (SSB)
dimensionful parameters of $N=1$ supersymmetric theories. In addition
it was found \cite{kkk1} that RGI SSB scalar masses in general Gauge-Yukawa
unified models satisfy a universal sum rule at one-loop, which was
subsequently extended first up to two-loops \cite{kkmz1} and then to
all-loops \cite{kkz}.

To be more specific, consider the superpotential given by
(\ref{supot}) along with the Lagrangian for SSB terms \bea -{\cal
  L}_{\rm SB} &=& \frac{1}{6} \,h^{ijk}\,\phi_i \phi_j \phi_k +
\frac{1}{2} \,b^{ij}\,\phi_i \phi_j\\
&+& \frac{1}{2} \,(m^2)^{j}_{i}\,\phi^{*\,i} \phi_j+ \frac{1}{2} \,M\,\lambda
\lambda+\mbox{H.c.},\nn \eea where the $\phi_i$ are the scalar parts of the
chiral superfields $\Phi_i$ , $\lambda$ are the gauginos and $M$ their
unified mass.  Since we would like to consider only finite theories
here, we assume that the one-loop $\beta$-function of the gauge coupling
$g$ vanishes.  We also assume that the reduction equations admit power
series solutions of the form $C^{ijk} = g\,\sum_{n=0}\,\rho^{ijk}_{(n)}
g^{2n}~.  $ According to the finiteness theorem of
ref.~\cite{lucchesi1}, the theory is then finite to all orders in
perturbation theory, if, among others, the one-loop anomalous
dimensions $\gamma_{i}^{j(1)}$ vanish.  The one- and two-loop finiteness
for $h^{ijk}$ can be achieved \cite{soft,zoup-jack1} by imposing the condition
\beq h^{ijk} = -M C^{ijk}+\dots =-M
\rho^{ijk}_{(0)}\,g+O(g^5)~.
\label{hY}
\eeq
In addition it was found \cite{kkmz1} that one and two-loop finiteness
requires that the following two-loop sum rule for the soft scalar
masses has to be satisfied \beq
\frac{(~m_{i}^{2}+m_{j}^{2}+m_{k}^{2}~)}{M M^{\dag}} = 1+\frac{g^2}{16
  \pi^2}\,\Delta^{(2)} +O(g^4)~,
\label{sumr} 
\eeq
where $\Delta^{(2)}$ is the two-loop correction,
\beq
\Delta^{(2)} =  -2\sum_{l} [(m^{2}_{l}/M M^{\dag})-(1/3)]~T(R_l),
\label{delta}
\eeq
which vanishes for the
universal choice \cite{zoup-jack1}.
Further, it was found \cite{jack4}  that the relation
\beq
h^{ijk} = -M (C^{ijk})'
\equiv -M \frac{d C^{ijk}(g)}{d \ln g}~,
\label{h}
\eeq 
among couplings is all-loop RGI. Moreover, the progress made using
the spurion technique leads to
all-loop relations among SSB $\beta$-functions
\cite{acta,hisano1,yamada1,kazakov1,jack4}, which allowed to find the
all-loop RGI sum rule \cite{kkz} in the
Novikov-Shifman-Vainstein-Zakharov scheme \cite{zoup-novikov1}. 

\section{Search for realistic FUTs based on product gauge groups}

Let us now examine the possibility of constructing realistic FUTs based
on product gauge groups. Consider the gauge group $SU(N)_1 \times SU(N)_2 \times
... \times SU(N)_k$  with $n_f$ copies of the supersymmetric multiplet
$(N,N^*,1,...,1) +  
(1,N,N^*,...,1) + ... + (N^*,1,1,...,N)$.  The one-loop $\beta$-function 
coefficient in the renormalization-group equation of each $SU(N)$ gauge 
coupling is simply given by
\begin{equation}
b = \left( -{11 \over 3} + {2 \over 3} \right) N + n_f \left( {2 \over 3} 
+ {1 \over 3} \right) \left( {1 \over 2} \right) 2 N = -3 N + n_f N.
\label{1}
\end{equation}
This means that $n_f = 3$ is a solution of the equation $b=0$,
independently of the values of $N$ and $k$.  Since $b=0$ is a necessary
condition for a finite field theory, the existence of three families of
quarks and leptons is natural in such models.  (This is true of course
only if the matter content is exactly as given above.  Other $SU(N)^k$
models exist with very different, and rather {\it ad hoc}, 
supermultiplet structure.  They are not included in our discussion.)

Next let us examine if this class of models can meet the obvious
requirements in every unified theory, namely (i) that it leads to the
SM or the MSSM at low energies, and (ii) that it predicts correctly $sin^2\theta_W$.

Let $N=3$ and $k=3$, then we have the well-known example of $SU(3)_C \times 
SU(3)_L \times SU(3)_R$ \cite{trini, su3^3}, with quarks transforming as
\begin{equation}
q = \pmatrix {d & u & h \cr d & u & h \cr d & u & h} \sim (3,3^*,1), ~~~ 
q^c = \pmatrix {d^c & d^c & d^c \cr u^c & u^c & u^c \cr h^c & h^c & h^c} 
\sim (3^*,1,3),
\label{2}
\end{equation}
and leptons transforming as
\begin{equation}
\lambda = \pmatrix {N & E^c & \nu \cr E & N^c & e \cr \nu^c & e^c & S} 
\sim (1,3,3^*).
\label{3}
\end{equation}
If we switch the first and third rows of $q^c$ together with the first and 
third columns of $\lambda$, we obtain the alternative left-right model first 
proposed in ref.~\cite{malr} in the context of superstring-inspired $E_6$. 
The breaking  down of $SU(3)^3$ to $SU(3)_C \times SU(2)_L \times SU(2)_R \times 
U(1)_{Y_L+Y_R}$ is achieved with the (3,3) entry of $\lambda$, and the 
further breaking of $SU(2)_R \times U(1)_{Y_L+Y_R}$ to $U(1)_Y$ with the 
(3,1) entry.

Let $N=3$ and $k=4$, then one example is the extension to include the chiral 
color of ref.~\cite{cc}.  Here $SU(3)_C$ is split up into 
$SU(3)_{CL}$ 
and $SU(3)_{CR}$.  This implies the existence of a neutral supermultiplet 
$\eta$ transforming as $(N^*,N)$ under these two groups.  Let 
$\langle \eta_{11} \rangle = \langle \eta_{22} \rangle = \langle \eta_{33} 
\rangle$, then $SU(3)_{CL} \times SU(3)_{CR}$ breaks back down to $SU(3)_C$ as 
desired.  However at this scale,
\begin{equation}
\alpha_s^{-1} = \alpha_{sL}^{-1} + \alpha_{sR}^{-1}
\label{4}
\end{equation}
and since $\alpha_{sL}$ and $\alpha_{sR}$ are to be unified with $\alpha_L$ 
and $\alpha_R$, the predicted value of $\alpha_s$ would be too small. Thus
this is not a candidate model of unification, unless the particle
content is also extended \cite{Perez-Lorenzana:1998ss}, in which case
finiteness would be lost.

Another possibility to consider is the quartification model of
ref.~\cite{bmw}.  Here unification is possible but only in the
nonsupersymmetric case.  In fact, $\sin^2 \theta_W = 1/3$ instead of the
canonical 3/8, and the unification scale of this model is only $4 \times
10^{11}$ GeV.

Let us now turn to the interesting $N=4$ and $k=3$ case \cite{su4^3}. The
obvious choice is $SU(4)_C \times SU(4)_L \times SU(4)_R$, where $SU(4)_C$ is
the Pati-Salam color gauge group \cite{ps74}.  In that case, the
quarks and leptons should transform as
\begin{equation}
f = \pmatrix {d & u & y & x \cr d & u & y & x \cr d & u & y & x \cr 
e & \nu & a & v} \sim (4,4^*,1), ~~~ f^c = \pmatrix {d^c & d^c & d^c & e^c 
\cr u^c & u^c & u^c & \nu^c \cr y^c & y^c & y^c & a^c \cr x^c & x^c & x^c & 
v^c} \sim (4^*,1,4).
\label{5}
\end{equation}
We see immediately that there have to be new heavy particles, i.e. the $x$ 
and $y$ quarks and the $v$ and $a$ leptons.  In addition, we need to consider 
the $h \sim (1,4,4^*)$ supermultiplet.

The unification of quarks and leptons within $SU(4)_C$ implies that their 
electric charge $Q$ should be given by
\begin{equation}
Q = {1 \over 2} (B-L) + I_{3L} + I_{3R}.
\label{6}
\end{equation}
However, the electric charges of the new heavy particles are not yet fixed.  
This is because $SU(4)$ contains two disjoint $SU(2)$ subgroups, one of 
which may be the usual $SU(2)_L$ or $SU(2)_R$, but the other is new. 
Therefore, another valid formula for $Q$ is given by 
\begin{equation}
Q = {1 \over 2} (B-L) + I_{3L} + I_{3R} + I'_{3L} + I'_{3R}.
\label{7}
\end{equation}
The quarks and leptons do not transform under $SU(2)'_L$ or $SU(2)'_R$, 
so their electric charges are not affected.

Using \Eq {6}, the charges of $f$, $f^c$, and $h$ are respectively
\begin{equation}
Q_f = \pmatrix {-1/3 & 2/3 & 1/6 & 1/6 \cr -1/3 & 2/3 & 1/6 & 1/6 \cr 
-1/3 & 2/3 & 1/6 & 1/6 \cr -1 & 0 & -1/2 & -1/2},
\label{8} 
\end{equation}

\begin{equation} 
Q_{f^c} = \pmatrix {1/3 & 1/3 & 1/3 & 1 \cr -2/3 & -2/3 & -2/3 & 0 \cr 
-1/6 & -1/6 & -1/6 & 1/2 \cr -1/6 & -1/6 & -1/6 & 1/2}, 
\label{9}
\end{equation}

\begin{equation} 
Q_h = \pmatrix {0 & 1 & 1/2 & 1/2 \cr -1 & 0 & -1/2 & -1/2 \cr 
-1/2 & 1/2 & 0 & 0 \cr -1/2 & 1/2 & 0 & 0}.
\label{10}
\end{equation}
Using \Eq {7}, they are instead
\begin{equation}
Q_f = \pmatrix {-1/3 & 2/3 & -1/3 & 2/3 \cr -1/3 & 2/3 & -1/3 & 2/3 \cr 
-1/3 & 2/3 & -1/3 & 2/3 \cr -1 & 0 & -1 & 0}, 
\label{11}
\end{equation}

\begin{equation} 
Q_{f^c} = \pmatrix {1/3 & 1/3 & 1/3 & -1 \cr -2/3 & -2/3 & -2/3 & 0 \cr 
1/3 & 1/3 & 1/3 & -1 \cr -2/3 & -2/3 & -2/3 & 0}, 
\label{12}
\end{equation}

\begin{equation} 
Q_h = \pmatrix {0 & 1 & 0 & 1 \cr -1 & 0 & -1 & 0 \cr 
0 & 1 & 0 & 1 \cr -1 & 0 & -1 & 0}.
\label{13}
\end{equation}

The two different charge assignments result in two different values of
\begin{equation} 
\sin^2 \theta_W = {\sum I_{3L}^2 \over \sum Q^2}
\label{14}
\end{equation}
at the unification scale.  Whereas it is equal to 3/8 as usual in the
former, it becomes 3/14 in the latter, which is not realistic.
Therefore we will discuss further only the case with the charge
assignments of Eqs. (\ref{8}--\ref{10}).

Since we do not admit any other matter supermultiplets, the symmetry 
breaking of $SU(4)_C \times SU(4)_L \times SU(4)_R$ must be achieved 
with the vacuum expectation values of the neutral scalar components 
of $f$, $f^c$, and $h$.  The best we can do is to let all the (3,3), 
(3,4), (4,3), and (4,4) entries of $h$ acquire vacuum expectation values, 
but then the $SU(4)^3$ symmetry is only broken down to $SU(4)_C \times 
SU(2)_L \times SU(2)_R \times U(1)_{L+R}$.  The extra unwanted $U(1)$ 
is necessarily present because in the decomposition of $SU(4)_L$ and 
$SU(4)_R$ to their $SU(2) \times SU(2) \times U(1)$ subgroups, the 
diagonal subgroup $U(1)_{L+R}$ cannot be broken by the representation 
$(1,4,4^*)$.  This problem persists even after the breaking of $SU(4)_C 
\times SU(2)_R$ by the (2,4) entry of $f^c$ to $SU(3)_C \times U(1)_Y$.

Since the unbroken $U(1)$ couples to all particles, including the known quarks 
and leptons, this model cannot be viable phenomenologically.  We are thus 
forced to conclude that $SU(4)_C \times SU(4)_L \times SU(4)_R$ with only the 
matter content of $f$, $f^c$, and $h$ is not a suitable candidate for 
a finite theory of all particles.

There is another important constraint for a realistic $SU(N)^k$ theory
of quarks and leptons, i.e. the proper masses must be obtained.
Excluding naturally nonrenormalizable terms in the superpotential,
then only bilinear and trilinear terms are allowed.  For the matter
content assumed here, it would be zero unless $N=3$ or $k=3$.  (We
exclude $N=2$ or $k=2$ for obvious reasons.)  If $N=3$, then we have
an invariant from the product of three $(3,3^*)$ supermultiplets.  If
$k=3$, then the invariant $(N,N^*,1)(1,N,N^*) (N^*,1,N)$ may be
formed.  Therefore, this discussion leads us naturally to the case
$SU(3)^3$.

\section{An all-loop $SU(3)^3$ FUT}

Here we will discuss in some detail the supersymmetric $SU(3)^3$ FUT
with three families.  In general a supersymmetric $E_6$ model in four
dimensions is easily obtained in compactifications of a
ten-dimensional $E_8$, appearing in the heterotic string, over
Calabi-Yau spaces \cite{green-book}.  Even more interesting is the
possibility to obtain softly broken supersymmetric $E_6$ type models
via coset space dimensional reduction \cite{forgacs,kubyshin} in
compactifications using non-symmetric coset spaces
\cite{Manousselis:2001re}.  Subsequently the $SU(3)^3$ can emerge
using the Wilson fluxes \cite{green-book,hosotani} in a
straightforward way.  What is less obvious to obtain is the
spontaneous symmetry breaking of $SU(3)^3$ down to the MSSM, however
it has been done already some time ago \cite{Lazarides:1993uw}.  It
requires introducing eight superfield of the type $(\lambda, q, q^c)$ and
five corresponding mirror superfields $(\bar{\lambda}, \bar{q},
\bar{q^c})$.  The details of this construction are given in
ref.~\cite{Lazarides:1993uw}.  Therefore what remains as an open
question is how to obtain the complete and detailed chain of breakings
of the ten-dimensional $E_8$ down to the four-dimensional MSSM, but
this is deeply related to the most fundamental problem of string
theory, and will not be addressed further here. For our purposes,
following \cite{Lazarides:1993uw}, we consider a supersymmetric
$SU(3)^3$ model with three families holding between the Planck $M_P$ and the
unification $M_{GUT}$ scales, which breaks spontaneously down to the
MSSM at $M_{GUT}$.

In order for all the gauge couplings to be equal at $M_{GUT}$, as is
suggested by the LEP results \cite{Amaldi:1991cn}, the cyclic symmetry $Z_3$ 
must be imposed, i.e.
\begin{equation}
q \to \lambda \to q^c \to q,
\label{15}
\end{equation}
where $q$ and $q^c$ are given in \Eq {2} and $\lambda$ in \Eq {3}.  Then,
according to the discussion in section 3, the first of the finiteness
conditions (\ref{2nd}) for one-loop finiteness, namely the vanishing
of the gauge $\beta$-function is satisfied.
 
Next let us consider the second condition, i.e. the vanishing of the
anomalous dimensions of all superfields.  To do that first we have to
write down the superpotential. If there is just one family, then there
are only two trilinear invariants, which can be constructed respecting
the symmetries of the theory, and therefore can be used in the
superpotential as follows
\begin{equation}
f ~Tr (\lambda q^c q) + {1 \over 6} f' ~\epsilon_{ijk} \epsilon_{abc} 
(\lambda_{ia} \lambda_{jb} \lambda_{kc} + q^c_{ia} q^c_{jb} q^c_{kc} + 
q_{ia} q_{jb} q_{kc}).
\label{16}
\end{equation}
In this case, the condition for vanishing anomalous dimension of each
superfield  is given by \cite{lucchesi1,ermushev1,finite1,kkmz1,acta} 
\begin{equation}
{1 \over 2} (3|f|^2 + 2|f'|^2) = 2 \left( {4 \over 3} g^2 \right)~.
\label{17}
\end{equation}
Quark and leptons obtain masses when the scalar parts of the
superfields $(\tilde N,\tilde N^c)$ obtain vacuum expectation values (vevs),
\begin{equation}
m_d = f \langle \tilde N \rangle, ~~ m_u = f \langle \tilde N^c \rangle, ~~ 
m_e = f' \langle \tilde N \rangle, ~~ m_\nu = f' \langle \tilde N^c \rangle.
\label{18}
\end{equation}
With three families, the most general superpotential contains 11 $f$
couplings, and 10 $f'$ couplings, subject to 9 conditions, due to the
vanishing of the anomalous dimensions of each superfield.  The
conditions are the following
\begin{equation}
\sum_{j,k} f_{ijk} (f_{ljk})^* + {2 \over 3} \sum_{j,k} f'_{ijk} (f'_{ljk})^* 
= {16 \over 9} g^2 \delta_{il},
\label{19}
\end{equation}
where
\begin{eqnarray}
&& f_{ijk} = f_{jki} = f_{kij}, \label{20}\\ 
&& f'_{ijk} = f'_{jki} = f'_{kij} = f'_{ikj} = f'_{kji} = f'_{jik}.
\label{21}
\end{eqnarray}
Quarks and leptons receive  masses when  the scalar part of the  
superfields $\tilde N_{1,2,3}$ and $\tilde N^c_{1,2,3}$ obtain vevs as follows
\begin{eqnarray}
&& ({\cal M}_d)_{ij} = \sum_k f_{kij} \langle \tilde N_k \rangle, ~~~ 
   ({\cal M}_u)_{ij} = \sum_k f_{kij} \langle \tilde N^c_k \rangle, \label{22} \\ 
&& ({\cal M}_e)_{ij} = \sum_k f'_{kij} \langle \tilde N_k \rangle, ~~~ 
   ({\cal M}_\nu)_{ij} = \sum_k f'_{kij} \langle \tilde N^c_k \rangle.
\label{23}
\end{eqnarray} 
Since we want to have, among other conditions, gauge coupling
unification, we will assume that the particle content of our finite
$SU(3)^3$ model below $M_U$ is that of the MSSM with three fermion
families, but only two Higgs doublets.  Therefore we have to choose
the linear combinations $\tilde N^c = \sum_i a_i \tilde N^c_i$ and
$\tilde N = \sum_i b_i \tilde N_i$ to play the role of the two Higgs
doublets, which will be responsible for the electroweak symmetry
breaking.  This can be done by choosing appropriately the masses in
the superpotential \cite{Leon:1985jm}, since they are not
constrained by the finiteness conditions.  Moreover, we choose that
the two Higgs doublets are predominately coupled to the third
generation.  Then these two Higgs doublets couple to the three
families differently, thus providing the freedom to understand 
their different masses and mixings.

Assuming for our purposes here that all $f'$ couplings
vanish\footnote{In supersymmetric theories this can always be done due
  to the non-renormalization theorem \cite{Wess:cp}, which guarantees
  that these terms will not appear radiatively.  In general this is
  not the case in the presence of supersymmetry breaking terms,
  however finiteness imposes tight conditions in this respect too.}
an isolated solution \Eq {19} is
\begin{equation}
f^2 = f^2_{111} = f^2_{222} = f^2_{333} = {16 \over 9} g^2.
\label{isosol}
\end{equation}
Hence we start at $M_{GUT}$ with different Yukawa couplings for all
the quarks 
\begin{eqnarray}
&& f_t = f a_3, ~~~ f_c = f a_2, ~~~ f_u = f a_1, \\ 
&& f_b = f b_3, ~~~ f_s = f b_2, ~~~ f_d = f b_1,
\end{eqnarray}
which is similar to the MSSM except that $f$ is fixed by finiteness at
$M_{GUT}$, and $a_3 \simeq 1$, $b_3 \simeq 1$, by construction, and therefore
we have that $f_t \simeq f_b \simeq f$ at $M_{GUT}$.  As for the lepton
masses, because all $f'$ couplings have been fixed to be zero at this
order, in principle they are expected to appear radiatively induced by
the scalar lepton masses appearing in the SSB sector of the theory.
Unfortunately though, due to the finiteness conditions (\ref{hY}) they
cannot appear radiatively and remain as a problem for further study.
On the other it should be stressed that we can certainly let $f'$ be
non-vanishing in \Eq{19} and thus introduce lepton masses in the
model.  Then the real price to be paid is basically aesthetic since
the model in turn becomes finite only up to two-loops since the
corresponding solution of \Eq{19} is not an isolated one any more.
However, given that the analysis we do in the next section takes into
account  RGEs up to two-loops, there is no practical cost in introducing
non-zero $f'$.  We include this possibility in our analysis in section
6.

Although we present the results of a more complete analysis in the next
section, we find instructive to  describe here the
situation concerning the top quark mass prediction at one-loop level
ignoring the SSB sector.  In this approximate analysis, we run the
MSSM renormalization group equations at one-loop, using our boundary
condition $f^2 = (16/9)g^2$ at the $M_{GUT}$ scale as follows
\begin{eqnarray}
8 \pi^2 (d g_3^2 /dt) &=& -3 g_3^4, \\ 
8 \pi^2 (d g_2^2 /dt) &=&  g_2^4, \\ 
8 \pi^2 (d g_1^2 /dt) &=& {33 \over 5} g_1^4, \\ 
8 \pi^2 (d f^2_t /dt) &=& f_t^2 \left( 6 f_t^2 + f_b^2 - {16 \over 3} g_3^2 - 
3 g_2^2 - {13 \over 15} g_1^2 \right), \label{difeq1}\\ 
8 \pi^2 (d f^2_b /dt) &=& f_b^2 \left( 6 f_b^2 + f_t^2 - {16 \over 3} g_3^2 - 
3 g_2^2 - {7 \over 15} g_1^2 \right). \label{difeq2}
\end{eqnarray}
The $g_i^2$s are easily solved as functions of $t=\ln (M_{GUT}/M)$:
\begin{eqnarray}
\alpha_3(M)^{-1} = \alpha_3(M_{GUT})^{-1} - (3/2\pi) \ln (M_{GUT}/M), \\ 
\alpha_2(M)^{-1} = \alpha_2(M_{GUT})^{-1} + (1/2\pi) \ln (M_{GUT}/M), \\ 
\alpha_1(M)^{-1} = \alpha_1(M_{GUT})^{-1} + (33/10\pi) \ln (M_{GUT}/M),
\end{eqnarray}
where $\alpha_i = g_i^2/4\pi$.  Using the MSSM boundary conditions from the 
unification of the gauge couplings at one-loop and the constraints of
the present model we have
\begin{eqnarray}
&& \alpha_i(M_{GUT}) = 0.0413, \\
&& \alpha_t(M_{GUT}) = \alpha_b(M_{GUT}) = (16/9) \alpha_i(M_{GUT}). 
\end{eqnarray}
Then we integrate the two differential equations (\ref{difeq1}) and
(\ref{difeq2}), from $t= \ln (M_{GUT}/M_{EW})$ to $t=0$, to determine
$f_t$ and $f_b$ at the electroweak scale $M_{EW}$. Then $m_t = f_t
v_u$ and $m_b = f_b v_d$, with $v_u$ and $V_d$ satisfying the
condition $v_u^2 + v_d^2 = v^2$, $v=174.3$ GeV.  Thus given $m_b$, we
can obtain $m_t$.

\section{Predictions and Conclusions}

The gauge symmetry $SU(3)^3$ is spontaneously broken down to the MSSM
at $M_{\rm GUT}$, and the finiteness conditions do not restrict the
renormalization properties at low energies. Therefore, below $M_{GUT}$
all couplings and masses of the theory run according to the RGEs of
the MSSM.  The remnants of the all-loop FUT $SU(3)^3$ are the boundary
conditions on the gauge and Yukawa couplings (\ref{isosol}), the
$h=-MC$ relation, and the soft scalar-mass sum rule (\ref{sumr})
at $M_{\rm GUT}$, which, when applied to the present model,
takes the form
\bea
m^2_{H_u} + m^2_{\tilde t^c} + m^2_{\tilde q} &=& M^2\\
m^2_{H_d} + m^2_{\tilde b^c} + m^2_{\tilde q} &=& M^2~.
\eea
Thus we examine
the evolution of these parameters according to their RGEs up to
two-loops for dimensionless parameters and at one-loop for
dimensionful ones imposing the corresponding boundary conditions. 
We further assume a unique supersymmetry breaking scale $M_{s}$
(defined as the average of the mass of the stops) and
therefore below that scale the effective theory is just the SM.

We consider two versions of the model:\\
{\bf I}) The all-loop finite one in which $f'$ vanishes and \Eq{isosol}
holds.\\
{\bf II}) A two-loop finite version, in which we keep $f'$
non-vanishing in \Eq{19},  and we use it to introduce the  lepton
masses.

The predictions for the top quark mass $m_t$ are $\sim$ 183 GeV for $\mu < 0 $ 
in model {\bf I}, whereas for model {\bf II} it is 176 - 179
GeV for $\mu  < 0$, and 170 -173 GeV for $\mu  > 0$. Recall that the
bottom quark mass $m_b$ is an input in {\bf FUT I} and $m_{\tau}$ in
{\bf FUT II}.

Comparing these
predictions with the most recent experimental value $ m_t^{exp} =
(178.0 \pm  4.3)$ GeV \cite{electroweak}, and recalling that the theoretical
values for $m_t$ may suffer from a correction of  $\sim 4 \%$
\cite{acta}, we see that they are consistent with the
experimental data. 

In the SSB sector, besides the constraints imposed by finiteness 
we further require\\ 
1) successful radiative electroweak symmetry breaking, and\\
2) $m_{\tilde\tau,\tilde b,\tilde t}^2 > 0$.\\
As an additional constraint, we take into account the  $BR(b \to s \gamma)$
\cite{zoup-bsg}.
We do not take into account, though, constraints coming from the muon
anomalous magnetic moment (g-2) in this work, which would exclude a small
region of the parameter space.  

Our numerical analysis shows the following results for the two models:
In the case of {\bf FUT I} it is possible to find regions of parameter
space which comply with all the above requirements both for the case
where we have universal boundary conditions ($m_i^2=m_j^2=m_k^2=
M^2/3$), and for the case where we apply the sum rule Eq.(\ref{sumr}).
In the case of universal boundary conditions and $\mu <0 $, $m_t \sim
183$ GeV, the Higgs mass is $\sim 131 -132$ GeV, $\tan \beta \sim 50-51$, and
the spectrum is rather heavy, the allowed region of parameter space
starting with an LSP which is a neutralino $m_{\chi^0} \sim 825$ GeV for a
value of $M \sim 1800$ GeV.  In the case the sum rule is applied we have
one more free parameter, which is $m_{\tilde q^c}=m_{\tilde q}$ at the
GUT scale.  In this case we obtain a $\tan \beta \sim 47-54$, and the Higgs
mass is $\sim 130-132$ GeV. The main difference between the universal
boundary conditions and the sum rule comes in the sparticle spectrum,
which can now start with an LSP at $m_{\chi^0} \sim 450$ GeV, for a
boundary condition of $M \sim 1800$ GeV.  In the case that $\mu > 0 $ we
do not find solutions which satisfy all the above requirements.

In the second version of the model {\bf FUT II}, we have the following
boundary conditions for the Yukawa couplings 
\bea
f^2 &=& r (16/9) g^2, \\
f'^2 &=& (1-r) (8/3) g^2.
\label{fprime}
\eea In this case, we do not have an all-loop finite model, since the
solution is a parametric one, but it is the price we pay to give
masses to the leptons.  As for the boundary conditions of the soft
scalars, we only have the universal case.  This is because, applying
the sum rule (\ref{sumr}) to the superpotential with $f' \neq 0$ implies
that $m_q^2=m_{q^c}^2=m_{H^{u,d}}^2= M^2/3$, which is again the
universal boundary condition.  For the numerical analysis we fix the
$m_{\tau}$ mass to obtain $m_t$ and $m_b$.  Taking $\mu <0$, and for
the experimentally allowed value of $m_b (m_b)=4.1-4.4$ GeV
\cite{PDG}, the value of $m_t$ goes from $\sim 176- 179$ GeV. In
this case $\tan\beta \sim 48 - 53$, and $m_H \sim 122-129$ GeV, with a
charged LSP $m_{\tilde\tau} \sim 400 - 1000$ GeV, depending directly on
the value of $M$, which varies from $\sim 1200 - 2200$ GeV in this
case.

Now for $\mu >0$, the value of $m_t$ compatible with the
experimentally allowed value of $m_b$ goes from $\sim 170-173$ GeV,
clearly the preferred value being the latter.  For this range of
values of $m_t$ we obtain $\tan\beta \sim 58 - 62$, and $m_H \sim
120-125$ GeV, also with a charged LSP $m_{\tilde\tau} \sim 300 - 600$
GeV, again depending directly on the value of $M$, which varies
from $\sim 1300 - 2000$ GeV.

We could go further and consider another version of the $SU(3)^3$
model.  For instance, if we impose global $SU(3)$ as a family symmetry
\cite{sun^3,kubo-priv}, then there is only one Yukawa coupling
in the superpotential, which leads to the following unique relation
among Yukawa and gauge couplings 
\beq 
f^2=\frac{8}{9}g^2 ~.  
\eeq 
However both ${\cal M}_u$ and ${\cal M}_d$ in \Eq{22} must now be
antisymmetric in family space, resulting in one zero and two equal
mass eigenvalues for each, which is not a realistic case.  Note
moreover, that the terms proportional to $f'$ in the superpotential
\Eq{16} are not allowed to appear in the cases of
refs.~\cite{su4^3,sun^3} unless $N=3$, and therefore they share the
problem of the {\bf FUT I} model, where we have chosen $f'=0$.
\section{Acknowledgements}

We acknowledge useful discussions with J. Erler, and use of some of
his results for the quark masses from his GAPP program
\cite{Erler:1999ug}.  We also acknowledge useful discussions with S.
Heinemeyer.  E.M. and G.Z. thank UNAM for its great hospitality during
a recent very productive visit.

\bibliographystyle{unsrt}

\end{document}